\def\year{2022}\relax
\documentclass[letterpaper]{article} 
\usepackage{aaai22}  
\usepackage{times}  
\usepackage{helvet}  
\usepackage{courier}  
\usepackage[hyphens]{url}  
\usepackage{graphicx} 
\usepackage{natbib}  
\usepackage{caption} 
\usepackage{balance}
\DeclareCaptionStyle{ruled}{labelfont=normalfont,labelsep=colon,strut=off} 
\usepackage{algorithm}
\usepackage{algorithmic}
\usepackage[dvipsnames]{xcolor}
\usepackage{mwe}
\usepackage{subcaption}
\usepackage{enumitem}
\usepackage{newfloat}
\usepackage{listings}
\floatstyle{ruled}
\newfloat{listing}{tb}{lst}{}
\floatname{listing}{Listing}

\title{Machine Learning-enhanced Efficient Spectroscopic Ellipsometry Modeling}
\author{
    Ayush Arunachalam \textsuperscript{\rm 1}\thanks{This work is 
    presented at the 1st Annual AAAI Workshop on AI to Accelerate Science and Engineering (AI2ASE) at the 36th AAAI Conference on Artificial Intelligence (AAAI 2022), and supported in part by the Semiconductor Research Corporation (GRC Task: 3026.001). \textit{Corresponding Author}: Ayush Arunachalam, \textit{Email}: arunachalamayush@utdallas.edu},
    S. Novia Berriel \textsuperscript{\rm 2},
    Parag Banerjee \textsuperscript{\rm 2},
    Kanad Basu  \textsuperscript{\rm 1}
}
\affiliations{
    \textsuperscript{\rm 1}University of Texas at Dallas, Richardson, TX 75080\\
    \textsuperscript{\rm 2}University of Central Florida, Orlando, FL 32816\\

}

\usepackage{bibentry}

\begin{document}

\maketitle

\begin{abstract}
Over the recent years, there has been an extensive adoption of Machine Learning (ML) in a plethora of real-world applications, ranging from computer vision to data mining and drug discovery. \textcolor{black}{In this paper, we utilize ML to facilitate efficient film fabrication, specifically Atomic Layer Deposition (ALD).} In order to make advances in ALD process development, which is utilized to generate thin films, and its subsequent accelerated adoption in industry, it is imperative to understand the underlying atomistic processes. Towards this end, \textit{in situ} techniques for monitoring film growth, such as Spectroscopic Ellipsometry (SE), have been proposed. However, \textit{in situ} SE is associated with complex hardware and, hence, is resource intensive.
To address these challenges, we propose an ML-based approach to expedite film thickness estimation. The proposed approach has tremendous implications of faster data acquisition, reduced hardware complexity and easier integration of spectroscopic ellipsometry for \textit{in situ} monitoring of film thickness deposition. Our experimental results involving SE of TiO\textsubscript{2} demonstrate that the proposed ML-based approach furnishes promising thickness prediction accuracy results of 88.76\% within $\pm$ 1.5 nm and 85.14\% within $\pm$ 0.5 nm intervals. Furthermore, we furnish accuracy results up to 98\% at lower thicknesses, which is a significant improvement over existing SE-based analysis, thereby making our solution a viable option for thickness estimation of ultrathin films.
\end{abstract}









\section{Introduction} 
\label{sec:intro}

Atomic Layer Deposition (ALD) is a method of film growth that makes use of two or more sequential self-limiting surface reactions. With ALD, one has the capability of producing thin films with a resolution in the order of angstroms. It is crucial to be able to measure the thickness of these films. Methods to determine film thickness include Scanning Electron Microscopy (SEM), Atomic Force Microscopy (AFM), X-ray Reflectometry (XRR), and Spectroscopic Ellipsometry (SE). Of these techniques, SE provides the opportunity for \textit{in situ} tracking of film thickness~\cite{RN1}. SE makes use of an elliptically polarized light source and a detector to measure the change in polarization of light incident on a sample. By using physics-based optical models to describe the substrate and film, thickness can be tracked in real-time. This real-time tracking can also provide insight into an important parameter in ALD, namely growth rate.

In order to execute the physics-based optical modeling, a modeling software must be used. This software and associated model libraries are supplied by the manufacturers of the SE equipment. The software makes use of a material model and attempts to fit that model to the raw polarization data obtained by the detector. By fitting this model to the raw data, properties like refractive index (\textit{n}), extinction coefficient (\textit{k}), and film thickness can be extracted. While this is an invaluable tool for ALD processes, it needs to be noted that SE is a technique in which the raw data must be fit to established models to obtain film properties.

Another possibility for analysis of SE data is the use of machine learning. \textit{In situ} SE produces a large volume of data \textemdash thousands of data points for a single process. If physics-based models are known \textit{a priori}, SE data and corresponding film thicknesses may be used to train an ML algorithm. Once trained, the algorithm could be used to extract thickness directly, without the need for a modeling step. Thus, the physics-based model serves only as a base on which an algorithm is trained.

\begin{figure}[t!]
\centering
\includegraphics[width=0.9\linewidth]{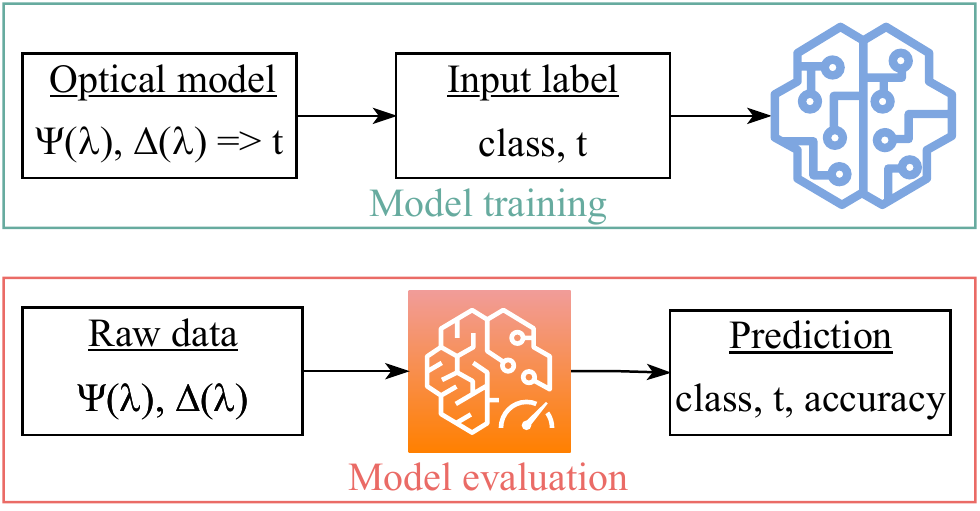}
\caption{Overview of the proposed ML film thickness estimation framework.}
 \label{fig:mlframework}
\vspace{-0.2in}
\end{figure}

Machine learning is becoming increasingly popular for identifying key descriptors for a given material~\cite{properties}. ML has been used to assist in analysis of Fourier transform infrared spectroscopy (FTIR)~\cite{ftir}, Raman~\cite{raman}, and X-ray diffraction (XRD)~\cite{xrd} for multi-component films with complex spectra. Correspondingly, a recent study used SE data along with transmission and reflection spectra to train an ML algorithm~\cite{liu2021machine}.
In this paper, we propose the use of ML algorithms to predict film thickness, given \textit{in situ} SE data of ALD TiO\textsubscript{2}. The physics-based model furnishes the thickness values at various $\Psi$ and $\Delta$, which constitutes the data used to train ML models. The trained models are subsequently utilized to determine thickness of TiO\textsubscript{2} films, as illustrated in Figure~\ref{fig:mlframework}, which illustrates the proposed approach.
The key contributions of this paper are:

\begin{itemize}[itemsep=0pt,parsep=0pt,leftmargin=*]

    \item This paper proposes an approach that incorporates machine learning to facilitate efficient estimation of film thickness in ALD. The proposed approach expedites film thickness estimation, with reliable prediction performance and consistent accuracy.
    

    
    \item Our extensive experiments on TiO\textsubscript{2} substrate demonstrate that the proposed approach can predict film thicknesses with an accuracy of 88.76\% within $\pm$1.5 nm and 85.14\% within $\pm$0.5 nm.

\end{itemize}

The rest of the paper is organized as follows. Section~\ref{sec:background} outlines the background and principal motivation for this research. The proposed ML framework is explained in detail in Section~\ref{sec:methodology}, while the experimental results are analyzed in Section~\ref{sec:results}. Finally, the paper is concluded in Section~\ref{sec:conclusion}.

\section{Background and Motivation}
\label{sec:background}

The use of ML to interpret characterization data is an efficient method to extract film properties \textemdash provided large databases are available for training. Prior research has investigated thin film characterization data involving SE using ML~\cite{liu2021machine}. This approach takes advantage of a large database of optical constants, namely refractive index (\textit{n}) and extinction coefficient (\textit{k}), and creates an iterative scheme for error minimization between the measured data and the one predicted by the ML model. Given the number of unknowns, the above approach relies on two measurement modalities – SE and reflectometry (and related transmission measurement) to define the problem in a complete manner. The transmission measurements pose a challenge for opaque substrates and it is not known how well the approach may work for films with interfaces and mixed compositions.

As an alternate to the above approach, we use ML-based algorithms to predict thickness solely from SE data. This data has been collected as part of \textit{in situ} experiments conducted on an ALD system with an SE attached to its chamber. The data is \textit{a priori} modeled for its thickness using a physics-based optical model. The optical modeling of the SE data provides us with the film thickness as a function of deposition time and serves as the training set for the ML algorithm. Here, we note that the quasi-continuous nature of the thickness data is important since it allows the ML algorithm to be trained across a wide range of thickness without having to resort to individual discrete experiments. In this context, the use of \textit{in situ} data is particularly advantageous.

\section{Proposed Methodology}
\label{sec:methodology}

\subsection{Dataset Creation}
\label{subsec:dataset}

\begin{figure}[t!]
\centering
\includegraphics[width=0.9\linewidth]{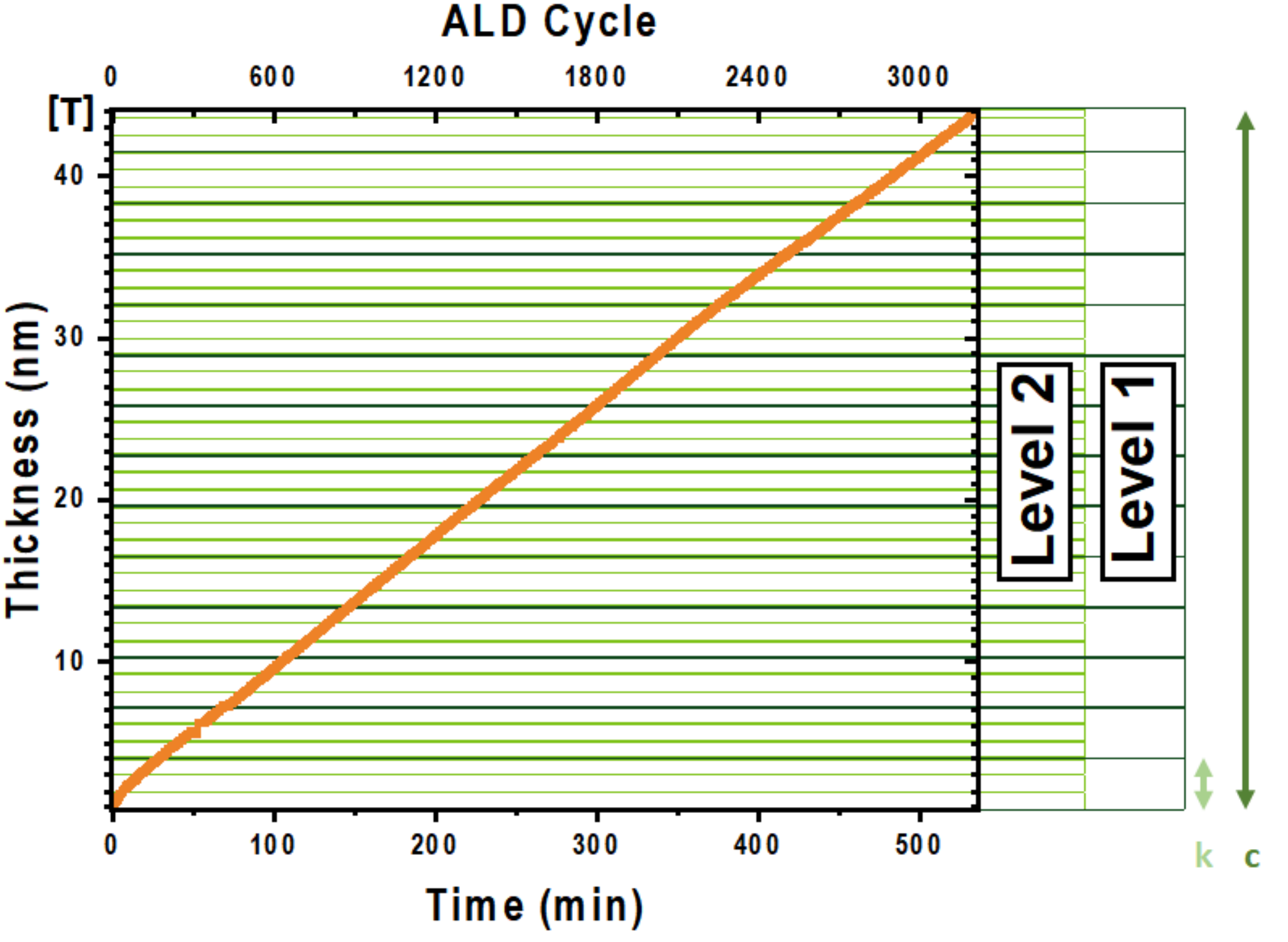}
\caption{Classes and subclasses constituting TiO\textsubscript{2} SE data.}
 \label{fig:tio2datamethodology}
\vspace{-0.1in}
\end{figure}

In this paper, we consider TiO\textsubscript{2} film thickness data, which as mentioned previously is quasi-continuous in nature, until 43 nm. First, this data is discretized by considering the integer values of thicknesses, as shown in Figure~\ref{fig:tio2datamethodology}. In addition to the variation of thickness with time, Figure~\ref{fig:tio2datamethodology} also depicts the different classes and subclasses comprising the classification data. Classification is achieved by defining a step-size of 0.2 nm and selecting films that range between \textit{x} and \textit{x+1} nm, where \textit{x} corresponds to a specific integer value of thickness. For example, films of thickness values 1 nm, 1.2 nm, 1.4 nm, 1.6 nm, and 1.8 nm are clubbed together and represented as a class comprising thickness of value 1 nm.
Subsequently, these integer film thicknesses are binned into $C$ classes, which have labels ranging from 0 to $C-1$, by employing a user-defined threshold of $k$ nm. The value of $C$ depends on the final thickness value of the film, $T$, as defined in equation~\ref{equ1}.
\begin{equation}
    \label{equ1}
   C = T / k
\end{equation}
Hence, each class comprises a maximum of $k$ subclasses (for $k$ thickness values). Depending on the value of $T$, the number of subclasses constituting the final class ranges between one and $k$. 

Next, we leverage this multi-class dataset to train various ML algorithms towards a multi-class classification task. While binary classification involves two classes, multi-class classification is associated with 
a range of classes, thereby aligning with our objective of classifying $\Psi$ and $\Delta$ into one of $C$ classes. Modeling such multi-class classification tasks primarily involves prediction of Multinoulli probability distribution for all data points; this, in our context, is equivalent to predicting the probability of a pair of $\Psi$ and $\Delta$ points belonging to each class.

\subsection{Application of Machine Learning}
\label{subsec:mlapproach}

The ML approach is evaluated on five algorithms, namely (1) k-Nearest Neighbors (kNN), (2) Random Forest (RF), (3) Decision Tree (DT), (4) Support Vector Machines (SVM), and (5) Logistic Regression (LR) using the curated multi-class data. These algorithms have been selected due to their proficiency and efficiency in multi-class classification tasks~\cite{sarker2021machine}. Furthermore, the aforementioned algorithms are robust, immune to overfitting during training, and require minimal to no prior knowledge about the distribution of input data.

Since the $\Psi$, $\Delta$ and thickness correlations have been established previously, it is convenient to train ML models. The training and test data are selected randomly, in a 95:5 split. First, a comparative analysis of the classification performance of all algorithms is performed, where the confidence in predicting a pair of $\Psi$ and $\Delta$ into one of $C$ classes is evaluated. For example, in our experiments, the final thickness ($T$) of the TiO\textsubscript{2} film is 43 nm. By using a threshold ($k$) of 3 nm, we obtain $C$ = 15 from Equation~\ref{equ1}. The ML model furnishing best accuracy is selected for subsequent analyses.

Next, this best model is assessed by incorporating a two-level evaluation, where the accuracy of classifying data points into different subclasses, in addition to different classes, is examined. The class- and subclass-level accuracy scores are labeled as level-1 accuracy and level-2 accuracy, respectively. The term level-1 accuracy describes the efficacy of classifying data points into one of $C$ classes and has a granularity of $k$ nm. As mentioned earlier, the values of $C$ and $k$ associated with our experiments are 15 and 3, respectively. Hence, each class comprises 3 subclasses, corresponding to 3 thicknesses. On the contrary, level-2 accuracy is associated with an increased granularity of 1 nm, which represents the effectiveness of predicting film thickness, down to 1 nm. To better comprehend model performance, we analyze classification accuracy scores at different thicknesses, in order to provide an insight into the variation of accuracy with classes and thicknesses.

\subsection{Downsampling}
\label{subsec: downsampling}

\begin{figure}[t!]
\centering
\includegraphics[width=\linewidth]{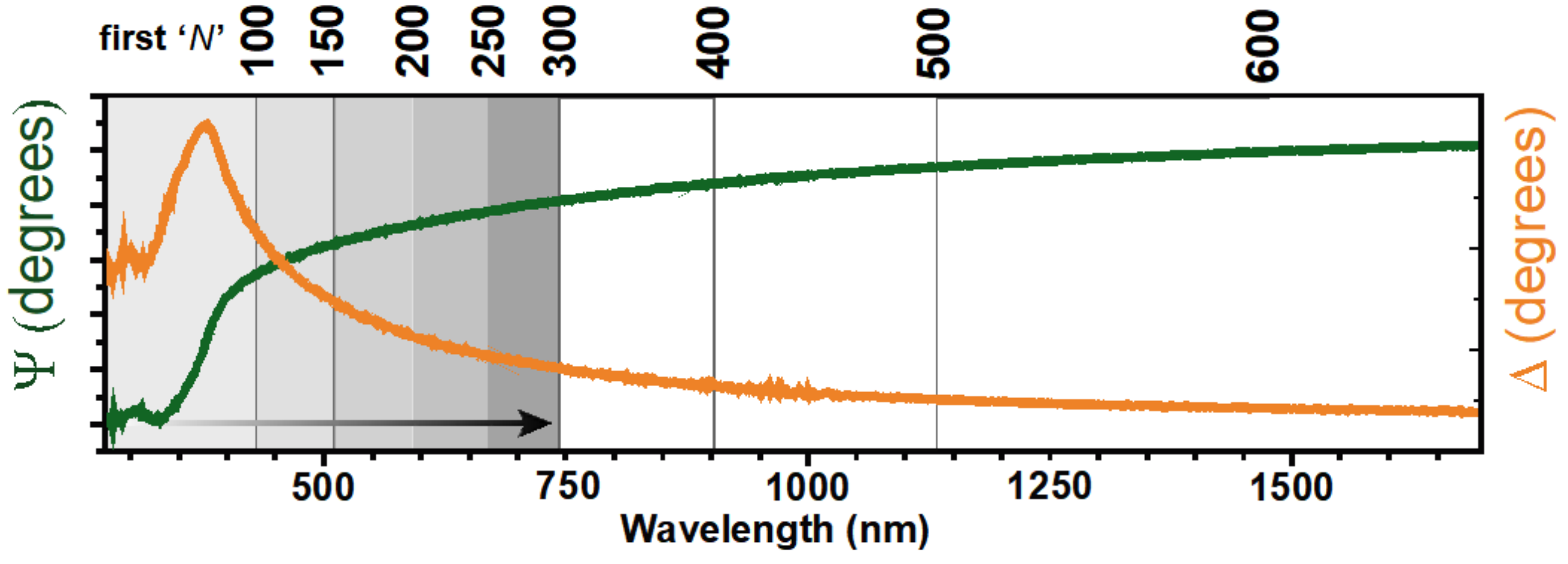}
\caption{Illustration of distribution of $\Psi$ (left-axis) and $\Delta$ (right axis) with respect to wavelength. Additionally, the first $N$ (where $N\epsilon \{100, 150, 200, 250, 300, 400, 500, 600\}$) data points constituting this distribution are also depicted here.}
 \label{fig:downsamplingfigure}
\vspace{-0.15in}
\end{figure}

The distribution of $\Psi$ and $\Delta$ values with respect to wavelength tends to contain a certain degree of redundancy, as depicted in Figure~\ref{fig:downsamplingfigure}. Hence, we incorporate downsampling in our approach, which entails removal of $\Psi$ and $\Delta$ values from our curated dataset. The goal is to understand how limited spectroscopic data might impact the classification performance of ML algorithms. The results of this analysis have huge implications of faster data acquisition, reduced hardware complexity (\emph{i.e.}, costs) and, subsequently, easier integration of SE for \textit{in situ} monitoring of thin film deposition processes, in general. 

Towards this end, we employ two downsampling techniques, namely random and selective downsampling. Random downsampling involves removal of random $\Psi$ and $\Delta$ values from the dataset. Selective downsampling, on the other hand, involves removal of specific $\Psi$ and $\Delta$ values. As illustrated in Figure~\ref{fig:downsamplingfigure}, distinct features, such as peaks and valleys, appear in a specific subset of the ellipsometry spectra, which corresponds to a specific number, $N$, of $\Psi$ and $\Delta$ samples. Hence, we select these $N$ samples from each class and subclass, while eliminating the rest. This analysis provides us an opportunity to explore the connection, if any, between specific $\Psi$ and $\Delta$ samples and prediction performance of ML models.

\section{Experiments}
\label{sec:results}

\subsection{Experimental Setup}
\label{sec:setup}

The substrate used for deposition in our experiments was silicon with native oxide. The substrate was ultrasonically cleaned in a mixture of isopropyl alcohol (IPA) and deionized H\textsubscript{2}O and dried with compressed air. The film was deposited using a Fiji Gen2 ALD System from Veeco®. The film grown was TiO\textsubscript{2}. The process consisted of alternate pulses of titanium tetraisopropoxide (TTIP 97\%, Sigma Aldrich) and deionized water.

The film thickness was tracked using an M-2000 spectroscopic ellipsometer from J. A. Woollam®. The light source and detector were mounted to the Fiji ALD system with 2.75$''$ conflat flanges and quartz windows. The windows are purged, protecting them from deposition, with 50 sscm of gaseous argon. The angle of incidence of the light on the sample was 69.5\textdegree. The wavelengths used for data acquisition ranged from 271 - 1688 nm, for a total of 661 different wavelengths per scan. The change in polarization from incident light to the reflected light was measured for the film. The quantities $\Psi$ and $\Delta$ are the raw data obtained from the ellipsometer, and they are representative of the change in polarization. A model was then used to fit the $\Psi$ and $\Delta$ as functions of wavelength to obtain film thickness.

The optical model for the film was created using the CompleteEase software from J.A. Woollam®. The model for the TiO\textsubscript{2} was a Cauchy-based model \cite{Woollam}. This model is part of a physics-based approach to estimate film thickness using SE data. To estimate film thickness via machine learning, the large amount of data generated by \textit{in situ} SE must be used to train an algorithm. The details of this process are given below.

The TiO\textsubscript{2} film thickness data ranges from 1 to 43 nm, \emph{i.e.}, the final thickness value of the film is $T$ = 43 nm. By employing a threshold $k$ = 3 nm, a classification dataset is curated to include 15 classes, as derived from Equation~\ref{equ1}. Hence, each class comprises three subclasses, except for the final one, which has one subclass. For example, class 0 consists of thickness values 1 nm, 2 nm, and 3 nm, while class 14 encompasses 43 nm. This dataset is used to train and evaluate various ML algorithms. Subsequently, a two-level classification strategy is adopted, as explained in Section~\ref{subsec:mlapproach}, where the first level, called level-1 classification, is associated with predicting one class (among the available 15) for a pair of $\Psi$ and $\Delta$ values, thereby having a granularity of 3 nm. On the other hand, the second level of classification, called level-2 classification, is associated with increased granularity of 1 nm, and furnishes the thickness value best describing a pair of $\Psi$ and $\Delta$ values.

\subsection{TiO\textsubscript{2} Ellipsometry Results}
\label{subsec:results}


\begin{figure}[t!]
\centering
\includegraphics[width=0.9\linewidth]{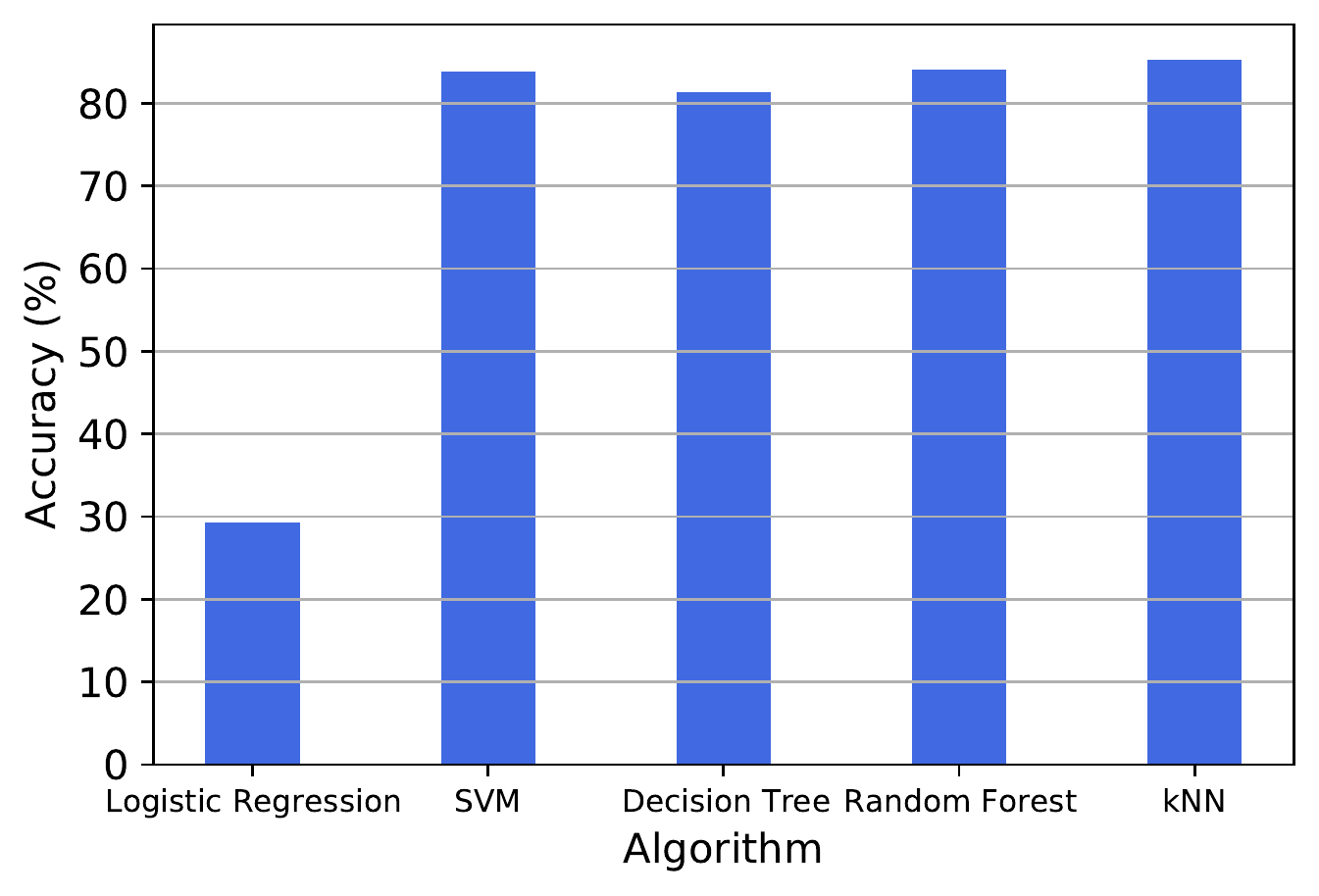}
\caption{Accuracy results of various ML classifiers.}
 \label{fig:tio2allclassifiers}
\vspace{-0.1in}
\end{figure}

\subsubsection{Comparative Analysis}
\label{subsubsec: analysisallalgos}
This case study compares the performance of the various ML algorithms, as demonstrated in Figure~\ref{fig:tio2allclassifiers}. The Y-axis (accuracy) illustrates the probability of classifying a data point into one of the 15 classes of the curated dataset to within $\pm$1.5 nm accuracy, \emph{i.e.}, level-1 classification. kNN furnishes the best accuracy score in this experiment, where each data point is classified with a confidence of 85.25\% using five neighbors~\cite{fix1989discriminatory, altman1992american}. The accuracy values reported by DT, SVM, and RF are 81.32\%, 83.84\%, and 84.1\%, respectively. The hyperparameters associated with SVM are a third-degree radial basis function and unit-value regularization coefficient. RF, which is an ensemble of decision trees, requires a minimum of two samples to split an internal node, similar to DT. We do not specify the maximum depth of the tree in either algorithm, thereby providing an opportunity for the tree to grow until all data samples are classified. In this experiment, we consider 40 trees in RF. LR, on the other hand, uses a regularization coefficient of value 1 and Limited-memory Broyden–Fletcher–Goldfarb–Shanno (lbfgs) optimization technique~\cite{fix1989discriminatory, altman1992american} to calculate the parameter weights that minimize the cost function.

LR performs the worst among all the models, producing an accuracy of 29.24\%. This can be attributed to LR being inept at solving non-linear problems owing to its linear decision surface, the requirement of having an exhaustive dataset, and its sensitivity of outliers in the dataset. Due to its low accuracy score, LR is not considered for future experiments. Additionally, SVM too is excluded from subsequent case studies for two reasons, (1) high training time and (2) sub-par classification performance. 
kNN, which furnishes the best results among the five algorithms, is incorporated into all our subsequent analyses on TiO\textsubscript{2} SE data.

\subsubsection{Multi-level Classification}
\label{subsubsec: multiclassification}
The second case study evaluates the performance of kNN with increasing granularity in the prediction of film thickness. These results assume importance from a process perspective since it is imperative to comprehend the uncertainty in accuracy (i.e., delta thickness), which is inherently present in ML algorithms, and the uncertainty that is propagated as a function of the true physical thickness of the film.

\begin{figure}[t!]
\centering
\includegraphics[width=0.9\linewidth]{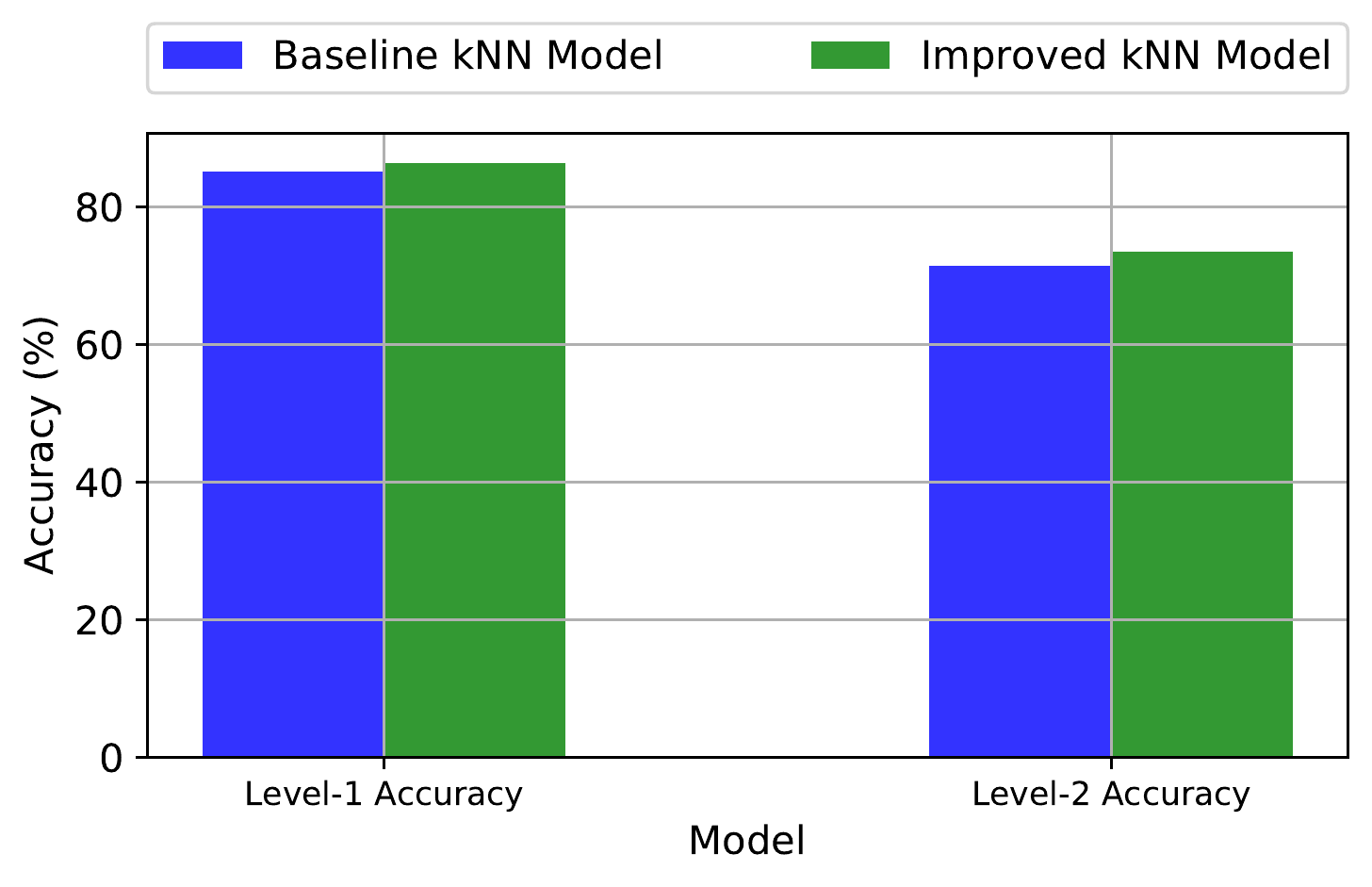}
\caption{Accuracy results of (a) baseline and (b) improved kNN model.}
\label{fig:tio2improvedmodels_three}
\end{figure}

The accuracy scores of kNN at both level-1 and level-2 are illustrated in Figure~\ref{fig:tio2improvedmodels_three}. We have updated the associated hyperparameters to bolster its classification performance. By increasing the number of neighbors from 5 to 21, we furnish an accuracy of 86.45\%, which is an enhancement of 1.2\% over the baseline model (85.25\% accuracy).



As illustrated in Figure~\ref{fig:tio2improvedmodels_three}, the baseline level-1 and level-2 accuracy scores of kNN on TiO\textsubscript{2} SE data, which are 85.25\% and 71.44\%, respectively, using five neighbors. The improved classification performance of the kNN model is also demonstrated in Figure~\ref{fig:tio2improvedmodels_three}, with level-1 accuracy now being 86.45\% and level-2 accuracy being 73.5\%, an increase of 1.2\% and 2.06\%, respectively.

\begin{figure}[t!]
\centering
\includegraphics[width=0.9\linewidth]{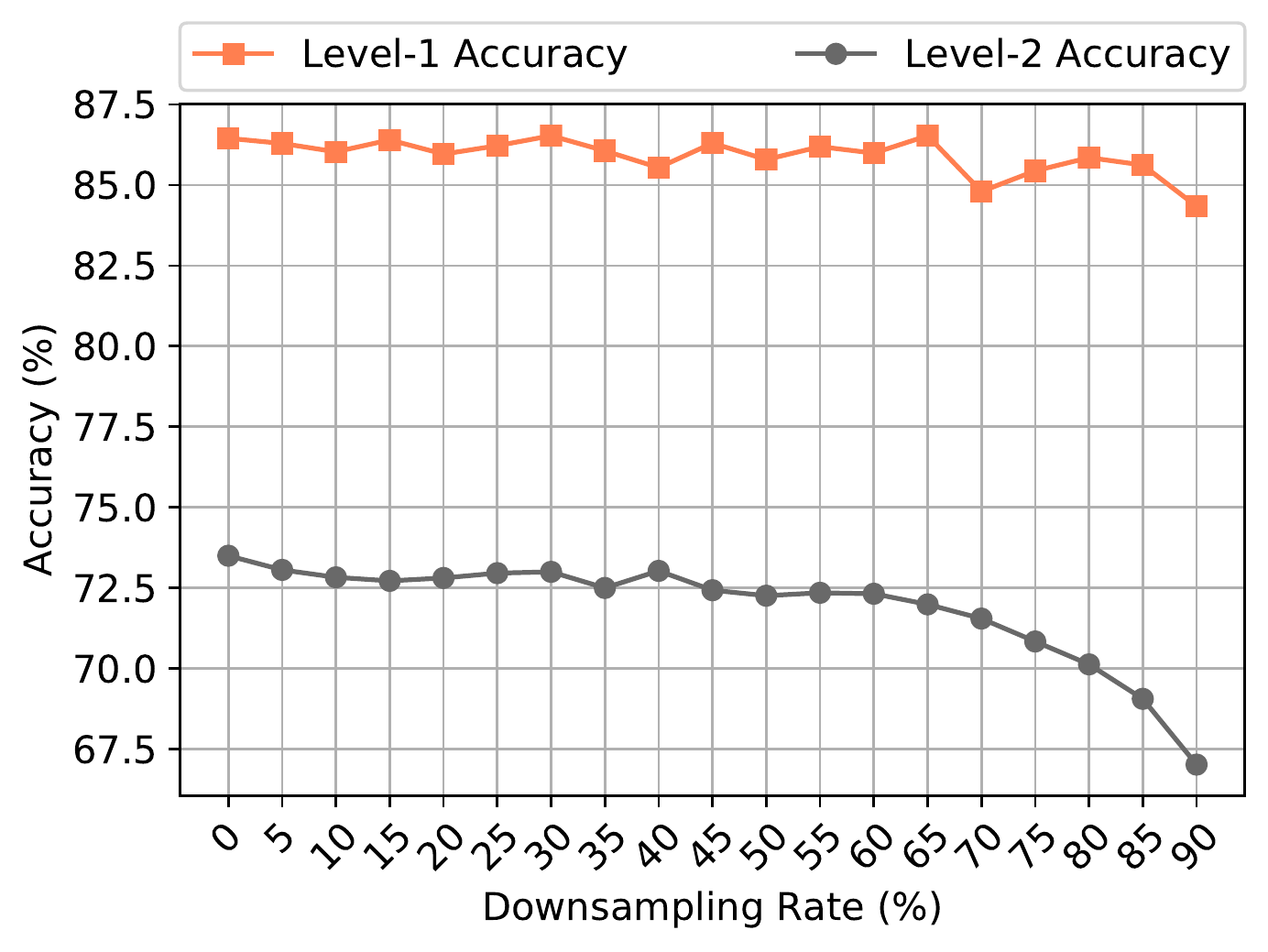}
\caption{Random downsampling results of kNN classifier for TiO\textsubscript{2} SE data.}
\label{fig:tio2dowsampling_acc}
\vspace{-0.15in}
\end{figure}

\subsubsection{Random Downsampling}
\label{subsubsec: randomdownsamplingresults}
The third case study describes the results of random downsampling of data, where random $\Psi$ and $\Delta$ values are removed from the curated dataset. In this experiment, we seek to understand the correlation between spectroscopic data and classification performance of ML models, specifically the impact of limited SE data on classification accuracy. The downsampling rate is varied between 0 and 90\% and the performance of the kNN algorithm is evaluated on this downsampled dataset. 
Figure~\ref{fig:tio2dowsampling_acc} depicts the variation of level-1 and level-2 accuracy values at different downsampling rates for TiO\textsubscript{2} data. Level-1 accuracy varies between 86.45\% and 85.62\% at 0\% and 85\% downsampling rates, respectively, furnishing a maximum accuracy drop of 0.83\%. Level-2 accuracy, on the other hand, varies from 73.5\% at 0\% downsampling to 67.02\% at 90\% downsampling, furnishing an accuracy drop of 6.48\%. 

From these results, we can infer that kNN is generally resilient to downsampling. This has major implications as it indicates that thickness prediction via ML can be effectively accurate at level-1 classification (to within ± 1.5 nm uncertainty) of a TiO\textsubscript{2} film $\leq43 nm$, even if 90\% of the spectral data is eliminated. For level-2 classification, however, negligible performance degradation is furnished until 65\% downsampling. Hence, the above results are indicative of possible redundancy in the $\Psi$ and $\Delta$ samples constituting SE dataset, which can be eliminated without inducing significant performance degradation. This subsequently motivates us to pursue selective downsampling and investigate the possible correlation between specific $\Psi$ and $\Delta$ samples and classification performance of ML models.

\begin{figure}[t!]
\centering
\includegraphics[width=0.9\linewidth]{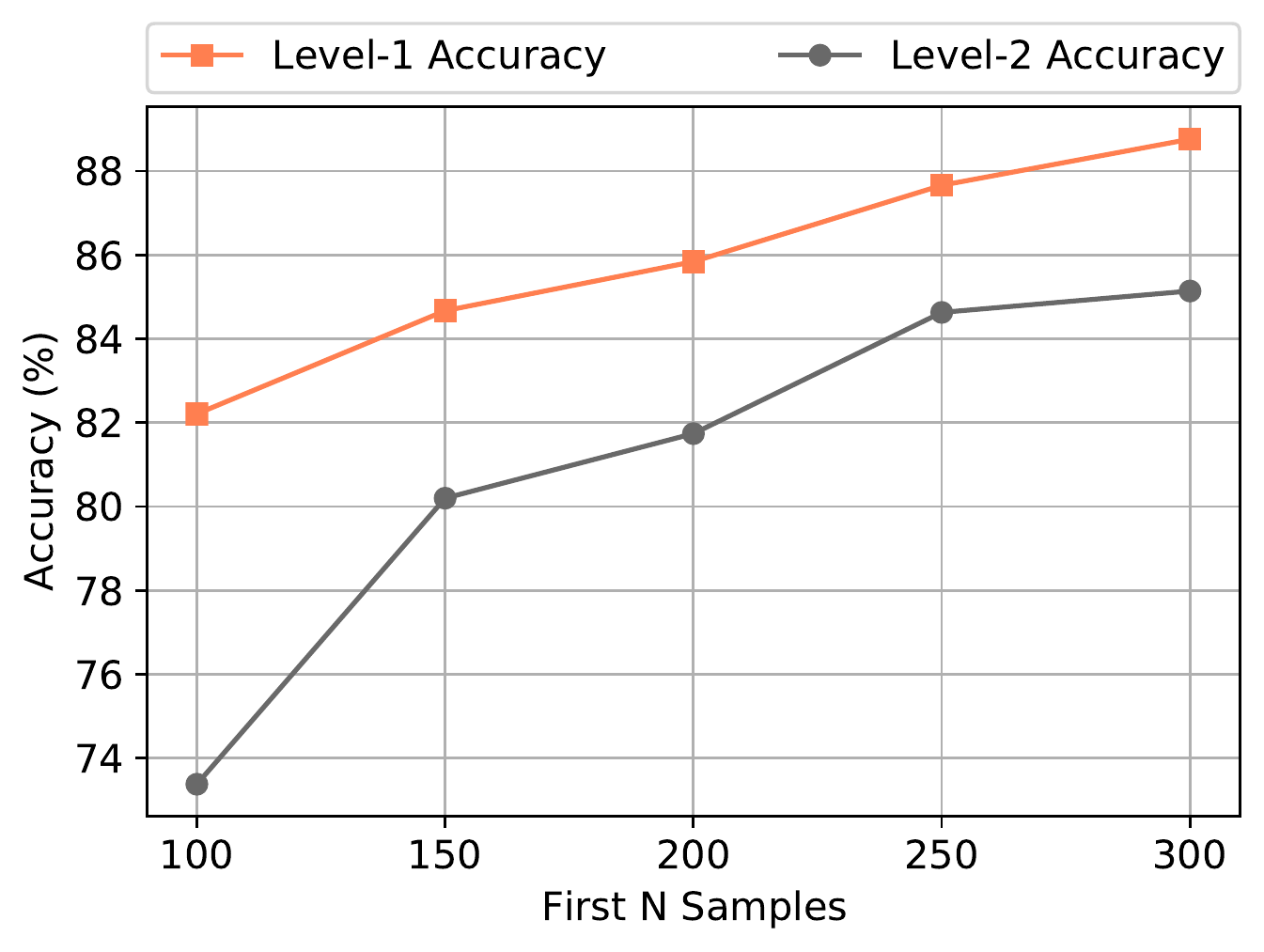}
\caption{Selective downsampling results of kNN classifier.}
\label{fig:tio2selectiveds_acc}
\vspace{-0.15in}
\end{figure}

\subsubsection{Selective Downsampling}
\label{subsubsec: selectivedownsamplingresults}
Since the ultimate objective involves ascertaining film thickness from minimum wavelength-based SE data samples, the fourth case study entails selective (non-random) downsampling of $\Psi$ and $\Delta$ values in the curated dataset. Towards this end, we select the first $N$ samples from each class and subclass, starting from 271 nm, while eliminating the rest, as shown in Figure~\ref{fig:downsamplingfigure}. The $\Psi$ and $\Delta$ values illustrated here correspond to a single spectrum, \emph{i.e.}, fixed thickness. Features such as peaks and valleys appear in the spectra for wavelengths $\leq750 nm$.

In this analysis, we consider $N$ ranging from 100 to 300, in step sizes of 50. This corresponds to a spectral window from 271 to 427 nm and 271 to 743 nm for the first 100 and 300 samples, respectively, in step sizes of 76 nm. Figure~\ref{fig:tio2selectiveds_acc} demonstrates the variation of level-1 and level-2 classification accuracy scores for TiO\textsubscript{2} data. While we obtain the lowest scores of 82.21\% at level-1 and 73.38\% at level-2 for the first 100 samples, improved classification performance is produced at higher values of N. The best level-1 accuracy and level-2 accuracy are 88.76\% and 85.14\%, which are furnished when the first 300 samples constitute the dataset. The baseline accuracy values, which are obtained without performing downsampling, are 86.45\% and 73.5\% at level-1 and level-2 classification, respectively, as shown in Figure~\ref{fig:tio2improvedmodels_three}. This approach of selective downsampling furnishes results exceeding the baseline accuracy values of kNN, by 2.31\% at level-1 and 11.64\% at level-2. 

\begin{figure}[t!]
\begin{subfigure}{0.5\columnwidth}
    \centering
    \includegraphics[width=\linewidth]{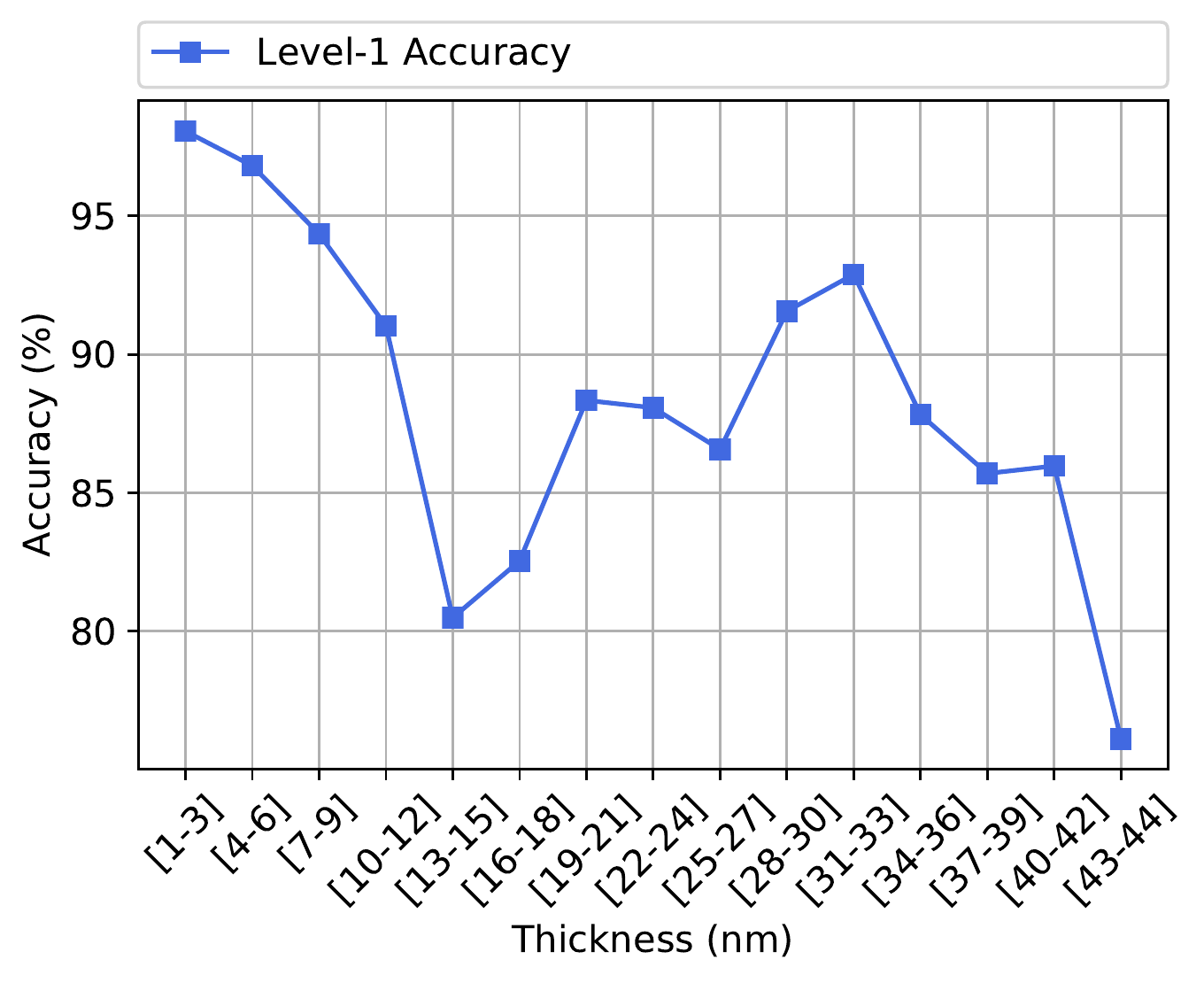}
    \caption{}
    \label{fig:tio2selectiveds_level1}
\end{subfigure}%
~
\begin{subfigure}{0.5\columnwidth}
    \centering
    \includegraphics[width=\linewidth]{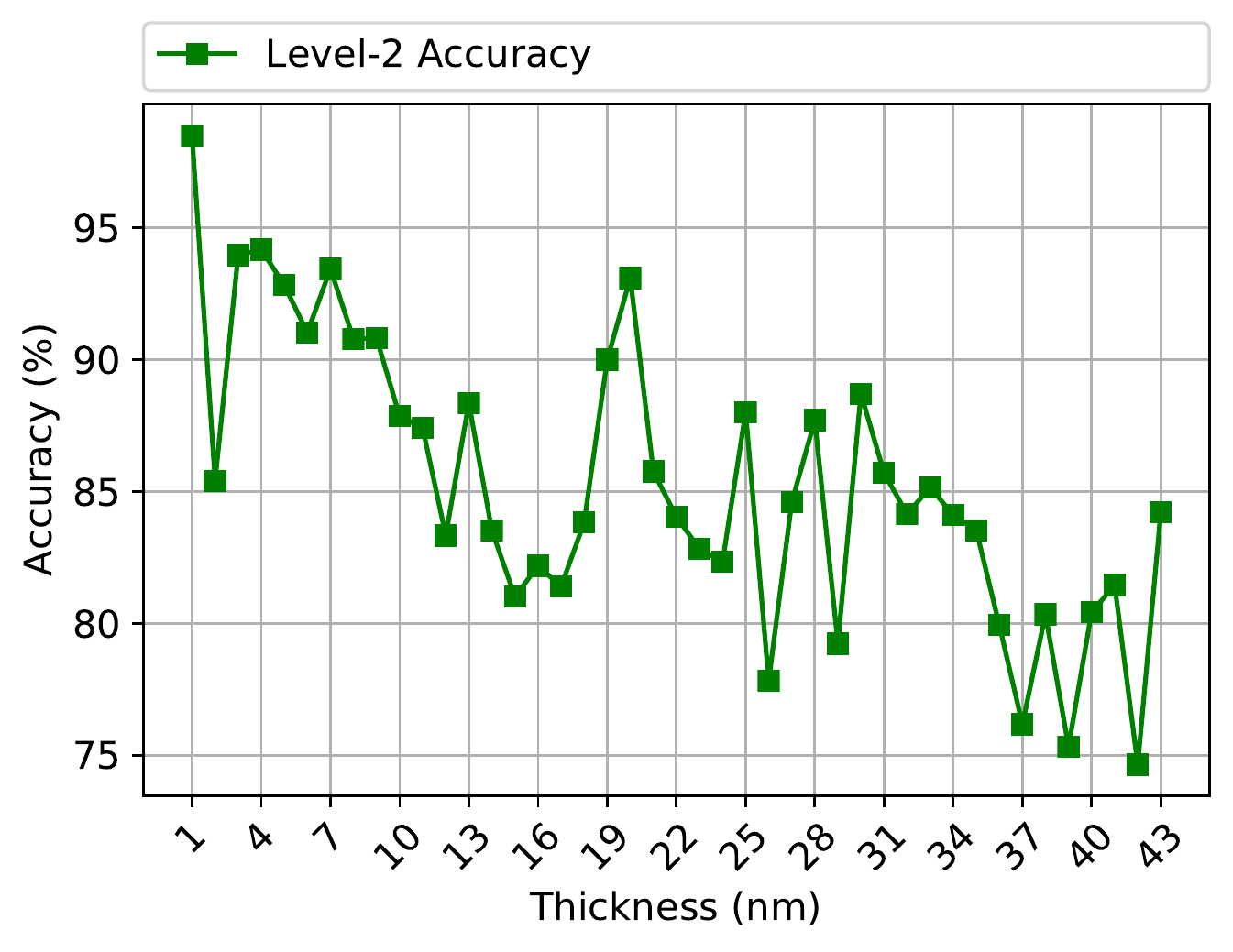}
    \caption{}
    \label{fig:tio2selectiveds_level2}
\end{subfigure}
\caption{Variation of accuracy of kNN classifier with (a) classes and (b) thickness for selective downsampling.}
\label{fig:tio2selectiveds_levels1-2}
\vspace{-0.15in}
\end{figure}

The variation of level-1 accuracy and level-2 accuracy with thickness for the first 300 samples are illustrated in Figure~\ref{fig:tio2selectiveds_level1} and Figure~\ref{fig:tio2selectiveds_level2}, respectively. While we obtain level-1 prediction accuracy of 98.1\% for a thickness in bin [1 - 3 nm], the accuracy degrades to 76.11\% for a thickness in bin [43 nm]. 
The difference in accuracy at different thicknesses is clearly illustrated by Figure~\ref{fig:tio2selectiveds_level2}; accuracy varies from 98.5\% at $t = 0$ to 84.21\% at $t = 43 nm$. From these graphs, it is evident that the classification performance of kNN degrades at higher thicknesses, which is attributed to the complex features associated with them. However, these results of high accuracy at lower thicknesses bode well for ML algorithms to predict thickness of ultrathin films ($\leq 10 nm$), \textbf{which is always a challenge for SE data analysis}.

Furthermore, the enhanced classification performance furnished by selective downsampling is accompanied by reduced data dimensionality; by selecting the first 300 out of the entire 661 data samples, we produce a dataset comprising a fraction (45.39\%) of the original SE data, thereby furnishing 54.61\% compression. Hence, such an approach facilitates an efficient, simple, and cost-effective thickness estimation procedure from SE data.

\section{Conclusion}
\label{sec:conclusion}

This paper proposes an ML approach to estimate thicknesses of films, fabricated via ALD, in an efficient manner. Complex hardware required to monitor and estimate film thickness in existing \textit{in situ} SE techniques renders them resource intensive. 
Our proposed approach incorporates ML to expedite film thickness estimation in a consistent and reliable manner. The $\Psi$ and $\Delta$ values obtained from physics-based optical models constitute our training dataset. The trained ML models are subsequently utilized to predict film thickness for a pair of random $\Psi$ and $\Delta$ values. Our experimental analyses demonstrate promising results of 88.76\% prediction accuracy within $\pm$1.5 nm and 85.14\% accuracy within $\pm$0.5 nm. Furthermore, we furnish prediction accuracy results up to 98\% at lower thicknesses, which significantly outperforms existing SE-based analysis. Hence, the proposed ML approach is a promising, alternate solution for thickness estimation of ultrathin films. The implications of the proposed approach are faster data acquisition, reduced hardware complexity associated with film estimation and, subsequently, easier integration of spectroscopic ellipsometry for \textit{in situ} monitoring of film thickness during deposition.

\balance

\bibliography{aaai22}

\end{document}